\documentclass[reprint,aps,pra,superscriptaddress,showpacs,amsmath,amssymb,floatfix]{revtex4-1}

\usepackage{txfonts}

\allowdisplaybreaks[4]

\usepackage{graphicx}
\usepackage{txfonts}
\usepackage{subfigure}
\usepackage{braket}
\usepackage{bm}

\usepackage{color}

\graphicspath{{./Figures/}}

\newcommand{\UTokyo}{
Department of Applied Physics, School of Engineering, \\
The University of Tokyo, 7-3-1 Hongo, Bunkyo-ku, Tokyo 113-8656, Japan}

\begin{document}

\title{Generation of optical Schr\"{o}dinger's cat states by generalized photon subtraction}

\author{Kan Takase}
%\email{takase@alice.t.u-tokyo.ac.jp}
\affiliation{\UTokyo}

\author{Jun-ichi Yoshikawa}
\affiliation{\UTokyo}

\author{Warit Asavanant}
\affiliation{\UTokyo}

\author{Mamoru Endo}
\affiliation{\UTokyo}

\author{Akira Furusawa}
\email{akiraf@ap.t.u-tokyo.ac.jp}
\affiliation{\UTokyo}

\date{\today}

\begin{abstract}
We propose a high-rate generation method of optical Schr\"{o}dinger's cat states. Thus far, photon subtraction from squeezed vacuum states has been a standard method in cat-state generation, but its constraints on experimental parameters limit the generation rate. In this paper, we consider the state generation by photon number measurement in one mode of arbitrary two-mode Gaussian states, which is a generalization of conventional photon subtraction, and derive the conditions to generate high-fidelity and large-amplitude cat states. Our method relaxes the constraints on experimental parameters, allowing us to optimize them and attain a high generation rate. Supposing realistic experimental conditions, the generation rate of cat states with large amplitudes ($|\alpha| \ge 2)$ can exceed megacounts per second, about $10^3$ to $10^6$ times better than typical rates of conventional photon subtraction. This rate would be improved further by the progress of related technologies. Ability to generate non-Gaussian states at a high rate is important in quantum computing using optical continuous variables, where scalable computing platforms have been demonstrated but preparation of non-Gaussian states of light remains as a challenging task. Our proposal reduces the difficulty of the state preparation and open a way for practical applications in quantum optics.
\end{abstract}

%\pacs{03.67.-a,42.50.Dv,42.50.Ex}
% % 03.67.-a : Quantum information
% % 03.67.Hk : Quantum communication
% % 42.50.Dv : Quantum state engineering and measurements
% % 42.50.Ex : Optical implementations of quantum information processing and transfer

\maketitle

\section{Introduction}
Quantum computers attract attentions as high-performance information processors, and implementations based on various physical systems have been extensively studied. Among these systems, an optical continuous-variables (CV) system is a promising candidate, where scalable quantum computing platforms have been already demonstrated \cite{Yokoyama2013,Larsen2019,Asavanant2019,Asavanant2020}. For practical use of the platforms, non-Gaussian states of light are an essential resource because they enable universal quantum computing on these platforms in a fault-tolerant way \cite{Lloyd1999,Gottesman2001,Ohliger2010}. Despite the importance of the non-Gaussian states, those stable supply is challenging because heralded generation of non-Gaussian states is probabilistic. With the current technology, the clock frequency of quantum computing would be strongly limited by the generation rate of non-Gaussian states rather than calculation platforms \cite{Asavanant2020}. Therefore, high-rate generation of non-Gaussian states is a key technology for optical CV quantum computing.

\begin{figure}[b]
	\begin{center}
		\includegraphics[bb= 0 0 1330 420,clip,width=0.47\textwidth]{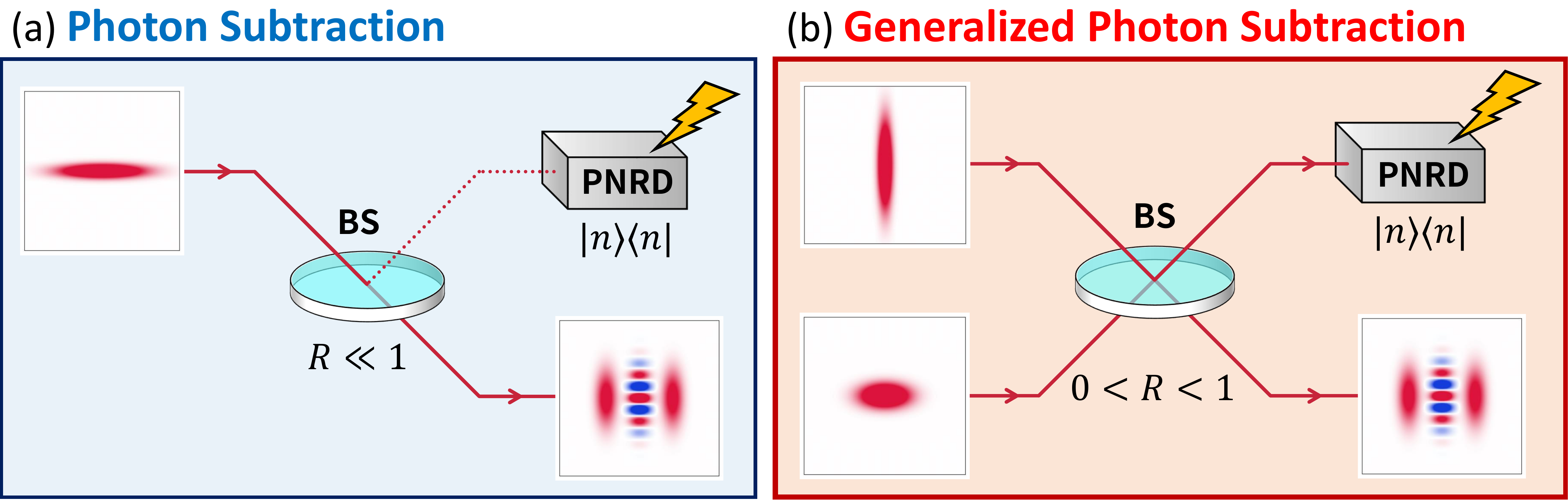}
	\end{center}
	\caption{(a) Photon subtraction method. A weak beam is tapped by a beam splitter (BS) from a squeezed vacuum. When a photon number resolving detector (PNRD) detects photons in the tapping channel, cat-like state are heralded. (b) Generalized photon subtraction (GPS). Two squeezed vacuum states squeezed in orthogonal directions interfere at a beam splitter. Photon number measurement heralds cat-like states.}
\label{Fig:ps_gps}
\end{figure}

Schr\"{o}dinger's cat states are typical non-Gaussian states of light, which are coherent-state superpositions given by $\ket{\alpha}\pm \ket{-\alpha}$. Large-amplitude cat states ($|\alpha| \ge 2$) can be utilized  as qubits of CV quantum computing \cite{Cochrane1999,Ralph2003} or resources for quantum error correction coding \cite{Vasconcelos2010,Weigand2018,Hastrup2020}. However, even in the best experiments, the generated optical cat states have the amplitudes $1.61 \le |\alpha| \le1.85$ \cite{Huang2015,Sychev2017,Ourjoumtsev2007,Ulanov2016,Gerrits2010,Takahashi2008}. This is mainly because the generation rate of the large-amplitude cat states is too low in conventional methods. A standard method of cat-state generation is the photon subtraction method shown in Fig.\ \ref{Fig:ps_gps}(a) \cite{Dakna1997,Takahashi2008,Ourjoumtsev2006,Neergaard-Nielsen2006,Wakui2007,Gerrits2010}. In this method, a squeezed vacuum is fed into a beam splitter whose reflectance is set $R\ll 1$ for beam tapping. By detecting photons in the tapping channel, we obtain cat-like states in the other output channel. To achieve $|\alpha| \ge2$, the number of detected photons should be $n\ge 4$ \cite{Dakna1997}. Such events are quite rare because the probability to detect $n$ photons has the order of $R^n$. Other proposed methods for cat-state generation \cite{Huang2015,Ourjoumtsev2007,Ulanov2016,Sychev2017} also suffer from a low generation rate. The generation rates in those methods are limited by the constraints on the power of input states \cite{Huang2015} or multiple conditioning processes \cite{Ourjoumtsev2007,Ulanov2016,Sychev2017}. Therefore, more efficient generation methods should be developed to generate large-amplitude cat states and utilize them in practical applications.

%%%%%%%%%%%%%%
%%%%%%%%%%%%%

In this paper, we propose a method for generation of optical Schr\"{o}dinger's cat states shown in Fig.\ \ref{Fig:ps_gps}(b), which we call ''generalized photon subtraction (GPS)''. In GPS, we consider performing photon number measurement in one mode of arbitrary two-mode Gaussian states and derive conditions to generate high-fidelity and large-amplitude cat states. GPS relaxes constraints on experimental parameters, thus we can avoid the undesirable condition $R\ll 1$, which limits the generation rate in conventional photon subtraction. Supposing realistic experimental conditions, the generation rate of cat states with $|\alpha| \ge 2$ can exceed megacounts per second (Mcps), about $10^3$ to $10^6$ times better than typical rates of conventional photon subtraction. This rate is clearly sufficient for state verification experiments, and, furthermore, is as fast as the system clock of current CV quantum information processors \cite{Asavanant2020}. In spite of the much improvement of the generation rate, GPS utilizes only two squeezed vacuum states, one beam splitter, and one photon number resolving detector (PNRD). Thus, the implementation of GPS is within reach of the current technology. Our proposal would reduce the difficulty in the state preparation and open a way for fault-tolerant CV quantum computing.

This paper is organized as follows; basics of optical cat states are given in Sec. \ref{2a}; we introduce GPS in Sec. \ref{2b}, and show the way of implementation in Sec. \ref{2c}; section \ref{2d} is devoted to discuss the validity of the condition of cat-state generation introduced in the previous sections; comparison of the generation rate with other methods is shown in Sec. \ref{3}; finally, we summarize our proposals in Sec. \ref{4}.

%%%%%%%%%%%%%%%%% END OF INTRODUCTION %%%%%%%%%%%%%%%%%%%%%%%%%%%%%%%%%%%%%

\section{GENERALIZED PHOTON SUBTRACTION}
\subsection{Optical Schr\"{o}dinger's cat states\label{2a}}
Optical Schr\"{o}dinger's cat states are often defined as superposition of coherent states with opposite phases. Coherent states are given by
\begin{eqnarray}\label{eq:cats}
\ket{\alpha}=\exp{(\alpha \hat{a}^{\dag}-\alpha^* \hat{a})}\ket{0},
\end{eqnarray}
where $\hat{a}$ and $\hat{a}^{\dag}$ are annihilation and creation operators and $\ket{0}$ is a vacuum state. $\hat{a}$ and $\hat{a}^{\dag}$ satisfy $[\hat{a},\hat{a}^{\dag}]=1$. Without loss of generality, we assume $\alpha \in \mathbb{R}, \alpha>0$ and define cat states as
\begin{eqnarray}\label{eq:cats}
\Ket{{\rm Cat}_{\alpha,k}} \equiv \frac{1}{N_{\alpha ,k}}\left[ \ket{\alpha} + (-1)^k \ket{-\alpha} \right],
\end{eqnarray}
where $N_{\alpha ,k} = \sqrt{2(1+(-1)^k\exp{(-2\alpha^2)})}$. Some previous works generated the squeezed cat states $\hat{S}(r)\Ket{{\rm Cat}_{\alpha,k}}$ \cite{Ourjoumtsev2007,Huang2015}, where $\hat{S}(r)=\exp{\left[ r\left( \hat{a}^{\dag  \ 2}-\hat{a}^2 \right)/2 \right]}\ \ (r \in \mathbb{R})$ is a squeezing operator. The squeezing operation can reduce the average photon number of cat states, and thus the squeezed cat states survive longer than usual cat states in lossy environment \cite{LeJeannic2018}. If we want to use the cat states that are not squeezed, we can unsqueeze them deterministically \cite{Miwa2014}.

Quadratures $\hat{x}=\left( \hat{a}+\hat{a}^{\dag} \right)/\sqrt{2}$ and $\hat{p}=\left( \hat{a}-\hat{a}^{\dag} \right)/\sqrt{2}i$ are useful tools to express quantum states of light. The quadratures satisfy a commutation relation $[\hat{x},\hat{p}]=i$. The wavefunctions of squeezed cat states are given by
\begin{eqnarray}
\Braket{x | \hat{S}(r) |{\rm Cat}_{\alpha,k}}
&\propto&  \mathrm{e}^{ -\frac{s^2}{2}\left(x-\sqrt{2}\alpha/s \right)^2 } + (-1)^k\mathrm{e}^{-\frac{s^2}{2}\left(x+\sqrt{2}\alpha/s \right)^2}, \label{eq:wavex} \\
\Braket{p|\hat{S}(r)|{\rm Cat}_{\alpha,k}} &\propto& \left( \mathrm{e}^{-i\sqrt{2}\alpha p/s} +(-1)^k \mathrm{e}^{i\sqrt{2}\alpha p/s} \right) \mathrm{e}^{-\frac{1}{2s^2}p^2}, \label{eq:wavep}
\end{eqnarray}
where $\ket{x}, \ket{p}$ are the eigenstates of $\hat{x}, \hat{p}$ and $s=\mathrm{e}^{r}$. The squeezing operator gives $\hat{S}^{\dag}(r)\hat{x}\hat{S}(r)=\hat{x}\mathrm{e}^{-r}$ and $\hat{S}^{\dag}(r)\hat{p}\hat{S}(r)=\hat{p}\mathrm{e}^{r}$, and thus $\hat{x}$ is squeezed when $r>0$. It is known that the function given in Eq.\ (\ref{eq:wavex}) with $\alpha=\sqrt{n}$ is well approximated as follows,
\begin{eqnarray}\label{eq:approxpsi}
\Braket{x \left| \hat{S}(r) \right|{\rm Cat}_{\sqrt{n},n}} \approx \Braket{x|\psi_{{\rm approx}}} \propto x^n\mathrm{e}^{-\frac{s^2}{4}x^2}.
\end{eqnarray}
The fidelity of $\hat{S}(r)\Ket{{\rm Cat}_{\sqrt{n},n}}$ and $\Ket{\psi_{{\rm approx}}}$ is $F_n \approx 1-0.03/n$ \cite{Ourjoumtsev2007}. In this paper, we propose a method to generate the cat-like state $\Ket{\psi_{{\rm approx}}}$.

%%%%%%%%%%%%%%%%%%%%%%%%%%%%%%%%%%%%%%%%%%%%%%%%%%%%%%%%%%%%%%%%%%%%%%%%%%%%%
\subsection{Generalized photon subtraction\label{2b}}
In conventional photon subtraction, a squeezed vacuum is fed into a beam splitter and photon number measurement is performed in a tapping mode to herald cat states (Fig.\ \ref{Fig:ps_gps}(a)). From a different perspective, the conventional photon subtraction consists of two parts: preparation of two-mode Gaussian states and non-Gaussian measurement on them. Note that in non-Gaussian state generation, either initial states or measurement should be non-Gaussian. It is desired that initial states are Gaussian because we can prepare them deterministically. The conventional photon subtraction only utilizes a small subspace of arbitrary two-mode Gaussian states due to its low degree of freedom, and thus there is room to find a more efficient generation method of cat states in a generalized situation. In this section, we consider the generation of cat states by photon number measurement on arbitrary two-mode Gaussian states, which we call generalized photon subtraction (GPS). Its experimental feasibility is discussed in Sec. \ref{2c}.

%%%%%%%%%%%%%%
\begin{figure}[t]
	\begin{center}
		\includegraphics[bb= 0 0 530 220,clip,width=0.45\textwidth]{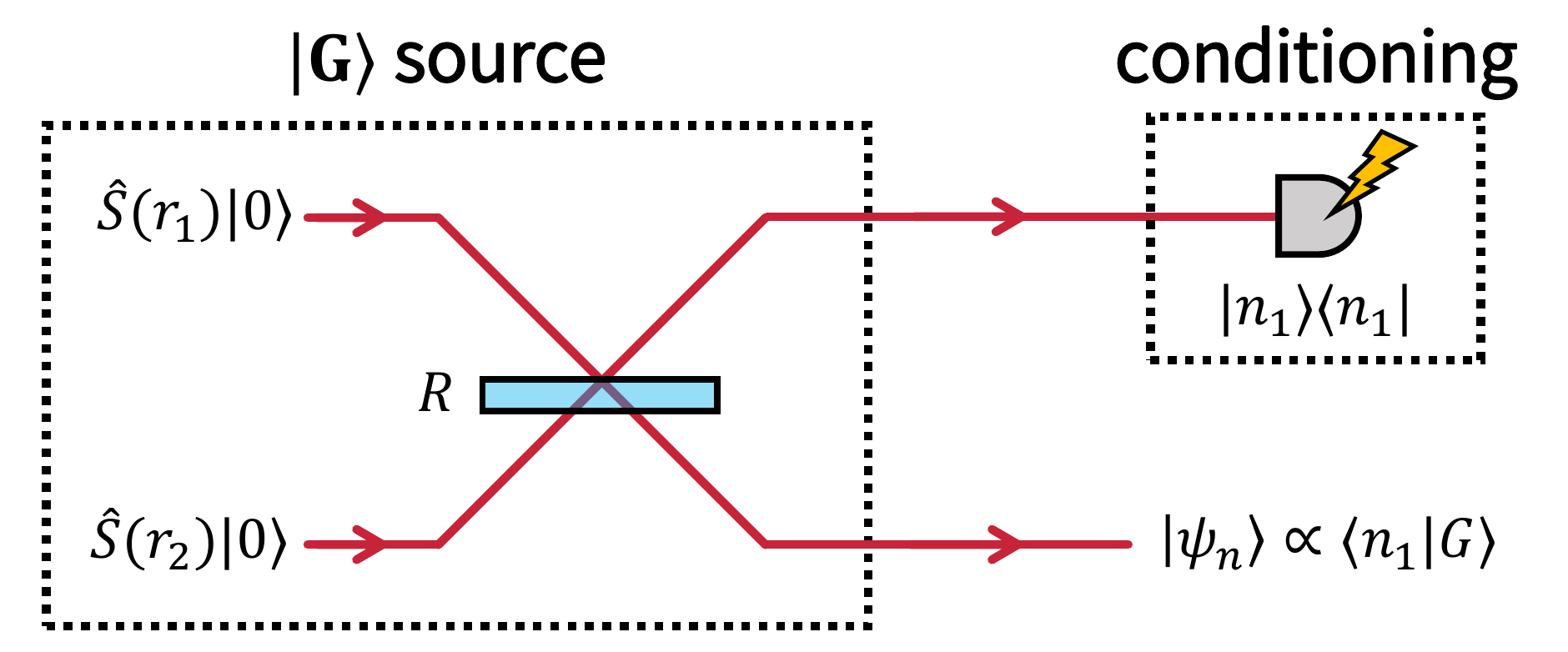}
	\end{center}
	\caption{Schematic of GPS. Firstly, a two-mode Gaussian state $\ket{G}$ is prepared. Detection of $n$ photons in one mode of $\ket{G}$ heralds $\ket{\psi_n}$ in the other mode. By preparing proper $\ket{G}$, the outcome state $\ket{\psi_n}$ approximates cat states well (Sec. \ref{2b}). Such $\ket{G}$ can be generated from two squeezed vacuum states and a beam splitter (Sec. \ref{2c}).}
\label{Fig:G}
\end{figure}
%%%%%%%%%%%%%

Firstly, we overview the process of state generation shown in Fig.\ \ref{Fig:G}. A two-mode Gaussian state $\Ket{G}$ is expressed by a complex Gaussian function $G(x_1,x_2)$ as
\begin{eqnarray}
\ket{G} = \iint dx_1 dx_2 \ G(x_1,x_2) \ket{x_1}\ket{x_2}.
\end{eqnarray}
When $n$ photons are measured in the mode 1, the state $\ket{G}$ is affected as follows,
\begin{eqnarray}
\braket{n_1|G} &=& \iint dx_1 dx_2 \ G(x_1,x_2) \braket{n_1|x_1}\ket{x_2} \nonumber \\
&=& \int dx_2 \left[ \int dx_1\ G (x_1,x_2)\phi_n (x_1) \right] \ket{x_2},
\end{eqnarray}
where $\phi_n(x)=\braket{x|n}$. Therefore, the unnormalized wavefunction of the outcome state is
\begin{eqnarray}\label{eq:wavefunction}
\Psi_n(x_2) =  \int dx_1\ G (x_1,x_2) \phi_n(x_1).
\end{eqnarray}
Thus, the bivariate function $G(x_1,x_2)$ linearly transforms $\phi_n(x_1)$ into $\Psi_n(x_2)$. The function $\Psi_n(x_2)$ is not normalized because the conditioning process is probabilistic. The probability $P(n)$ to detect $n$ photons and normalized wavefunction $\psi_n(x_2)$ are given by
\begin{eqnarray}
P(n) &=&  \int dx_2 \left| \Psi_n(x_2) \right|^2, \label{eq:pn} \\
\psi_n(x_2) &=& \frac{1}{\sqrt{P(n)}}\Psi_n(x_2).
\end{eqnarray}

Secondly, let us consider the conditions imposed to $G(x_1,x_2)$ in cat-state generation. $G(x_1,x_2)$ is given by
\begin{eqnarray}
&&G(x_1,x_2) = \frac{|\sigma |^{\frac{1}{4}}}{\sqrt{\pi}} \exp{\left[ -\frac{1}{2} (\bm{x}-\bm{\mu})^T \sigma (\bm{x}-\bm{\mu})-i\bm{x}^T\bm{\nu} \right]}, \\
&&\bm{x} = \left(
    \begin{array}{c}
      x_1  \\
      x_2
    \end{array}
  \right) \ , \ \bm{\mu} = \left(
    \begin{array}{c}
      \mu_1 \\
      \mu_2
    \end{array}
  \right) \ , \ 
\bm{\nu} = \left(
    \begin{array}{c}
      \nu_1  \\
      \nu_2
    \end{array}
  \right) \ , \ \sigma = \left(
    \begin{array}{cc}
      \sigma_{11} & \sigma_{12} \\
      \sigma_{21} & \sigma_{22}
    \end{array}
  \right), \nonumber
\end{eqnarray}
where $|M|$ denotes determinant of a matrix $M$. $\sigma$ satisfies $\sigma =\sigma^T$ and its elements are complex numbers in general. $\sigma$ becomes a real matrix when quadratures $x_1,x_2$ and $p_1,p_2$ are uncorrelated. $\bm{\mu}$ and $\bm{\nu}$ denote the displacement of $\ket{G}$ about $x$ and $p$. The wavefunction of Fock state $\ket{n}$ is given by
\begin{eqnarray}
\phi_n(x) = \frac{1}{\pi^{1/4}\sqrt{2^nn!}}H_n(x)\ \mathrm{e}^{-\frac{1}{2}x^2},
\end{eqnarray}
where $H_n(x)$ is a $n$-th order Hermite polynomial $H_n(x)=(-1)^n\mathrm{e}^{x^2} \frac{d^n}{dx^n}\mathrm{e}^{-x^2}$. Our target states are the cat states with $\alpha \in \mathbb{R}$, and thus $\Psi_n(x_2)$ should be an even or odd real function with non-Gaussian profile. From the symmetry $\phi_n(-x)=(-1)^n\phi_n(x)$, it is sufficient to assume the case where $\bm{\mu}=\bm{\nu}=\bm{0}$ and $\sigma$ is a real positive-definite matrix. From Eq.\ (\ref{eq:approxpsi}), we expect $\Psi_n(x)\propto x^n \mathrm{e}^{-\frac{s^2}{4}x^2}$. To obtain this function, we utilize a relation given by
\begin{eqnarray}\label{eq:vac_conv}
\left(\phi_0*\phi_n\right)(x) = \frac{1}{\sqrt{2^nn!}}x^n\ \mathrm{e}^{-\frac{1}{4}x^2},
\end{eqnarray}
where $\left( f*g \right) (x)$ denotes the convolution of $f(x)$ and $g(x)$. This equation is derived from an integral formula,
\begin{eqnarray}
\int_{-\infty}^{\infty} dy\ H_n(y)\ \mathrm{e}^{-(y-x)^2} = \sqrt{\pi} (2x)^n.
\end{eqnarray}
We transform $G(x_1,x_2)$ so that we can use Eq.\ (\ref{eq:vac_conv}) in the calculation of Eq.\ (\ref{eq:wavefunction}),
\begin{eqnarray}\label{eq:g}
&G&(x_1,x_2) \nonumber \\
&=& \frac{|\sigma |^{\frac{1}{4}}}{\sqrt{\pi}}\exp{\left[ -\frac{1}{2} \left( \sigma_{11}x_1^2 +2\sigma_{12}x_1x_2 + \sigma_{22}x_2^2 \right) \right]} \nonumber \\
&=& \frac{|\sigma |^{\frac{1}{4}}}{\sqrt{\pi}}\exp{\left[ -\frac{\left| \sigma \right|}{2\sigma_{11}}x_2^2 \right]} 
 \exp{\left[ -\frac{1}{2}\sigma_{11} \left( x_1+\frac{\sigma_{12}}{\sigma_{11}} x_2 \right)^2 \right]} \nonumber \\
&=& |\sigma |^{\frac{1}{4}}\ 
\phi_0 \left( \sqrt{\frac{\left| \sigma \right|}{\sigma_{11}}} x_2 \right) 
 \phi_0 \left( -\sqrt{\sigma_{11}} \left( x_1+\frac{\sigma_{12}}{\sigma_{11}} x_2 \right) \right).
\end{eqnarray}
We can use Eq.\ (\ref{eq:vac_conv}) when the following relations are satisfied,
\begin{eqnarray}\label{cons}
\sigma_{11} = 1\ ,\ \sigma_{12} \neq 0.
\end{eqnarray}
The latter condition is obvious because $\sigma_{12}=0$ means $x_1$ and $x_2$ are independent, hence photon number measurement does not affect the state in the other mode. When Eq.\ (\ref{cons}) is satisfied, we get
\begin{eqnarray}\label{eq:unnormwf}
\Psi_n(x_2)
&=& |\sigma |^{\frac{1}{4}} \phi_0 \left( \sqrt{\left| \sigma \right|}x_2 \right) \int dx_1\  \phi_0\left( -x_1-\sigma_{12}x_2 \right)\phi_n(x_1) \nonumber \\
&=& |\sigma |^{\frac{1}{4}} \phi_0 \left( \sqrt{\left| \sigma \right|}x_2 \right)
\left( \phi_0 *\phi_n \right)(- \sigma_{12} x_2) \nonumber \\
&=& \left( \frac{|\sigma |}{\pi} \right)^{\frac{1}{4}}\frac{(- \sigma_{12})^n}{\sqrt{2^nn!}}\ 
x_2^n \exp{\left(-\frac{\left| \sigma \right|+\sigma_{22}}{4} x_2^2 \right)}.
\end{eqnarray}
From Eq.\ (\ref{eq:approxpsi}), the outcome state satisfies
\begin{eqnarray}\label{eq:outcome}
\ket{\psi_n} \approx \hat{S}(r_c)\Ket{{\rm Cat}_{\sqrt{n},n}}\ ,\ \mathrm{e}^{2r_c} = \left| \sigma \right|+\sigma_{22}.
\end{eqnarray}
Therefore, the detection of $n$ photons heralds cat-like states with the amplitude $\alpha =\sqrt{n}$. As we mentioned in Sec. \ref{2a}, the fidelity of $\ket{\psi_n}$ and $\hat{S}(r_c)\Ket{{\rm Cat}_{\sqrt{n},n}}$ is $F_n \approx 1-0.03/n$ \cite{Ourjoumtsev2007}.

Summarizing the above, we introduced GPS as photon number measurement on arbitrary two-mode Gaussian states. In the conditions of $\sigma_{11}=1$ and $\sigma_{12}\neq 0$, the outcome states approximate cat states well as shown in Eq.\ (\ref{eq:outcome}). The dependence of the outcome states on $\sigma_{11}$ is discussed in Sec. \ref{2d}.

%%%%%%%%%%%%%%
\begin{figure*}[t]
	\begin{center}
		\includegraphics[bb= 0 0 1120 950,clip,width=0.9\textwidth]{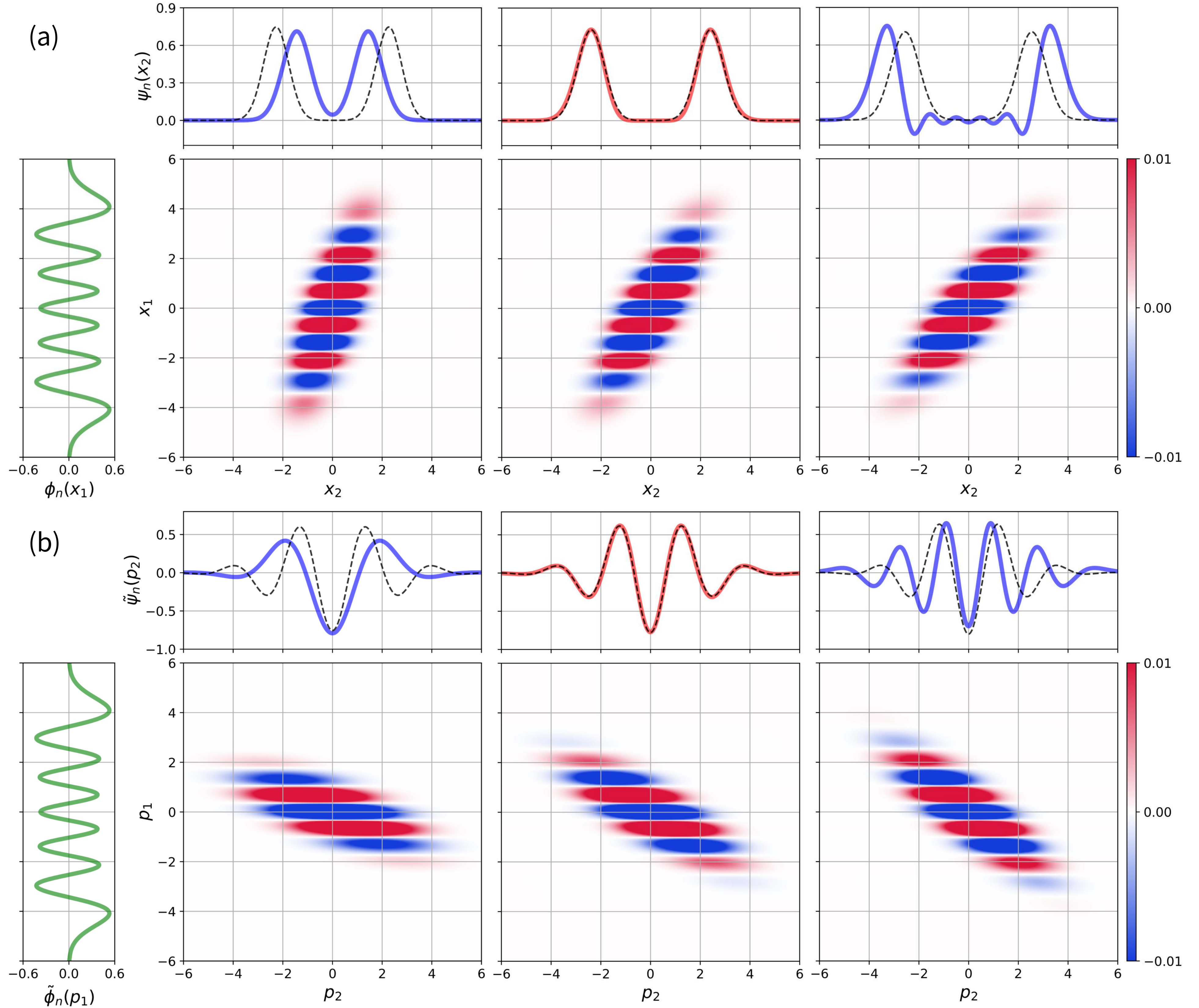}
	\end{center}
	\caption{(a) The plots of the functions $\phi_n(x_1)$, $\phi_n(x_1)G(x_1,x_2)$, and $\psi_n(x_2)$ in the case of $r_1 = -r_2 = 0.576$ (5 dB squeezing) and $n=10$. From the left,  $\sigma_{11}=0.6\ (R=0.10),\ \sigma_{11}=1\ (R=0.24),$ and $\sigma_{11}=1.4\ (R=0.38)$. The wavefunction of the target state in the case of $\sigma_{11}=1$ is shown in black broken lines. When $\sigma_{11}=1$, the fidelity of $\ket{\psi_n}$ and the target state is $F_{10}\approx 0.997$. (b) Similar plots about $p$.}
\label{Fig:main1}
\end{figure*}
%%%%%%%%%%%%%

%%%%%%%%%%%%%%%%%%%%%%%%%%%%%%%%%%%%%%%%%%%%%%%%%%%%%%%%%%%%%%%%%%%%%%%%%%%%%
\subsection{Preparation of two-mode Gaussian states \label{2c}}

The two-mode Gaussian state $\ket{G}$ used in GPS is generated from the interference of two squeezed vacuum states at a beam splitter as shown in Fig.\ \ref{Fig:G}. When the quadratures $x_1,x_2$ and $p_1,p_2$ are uncorrelated, the matrix $\sigma^{-1}$ is equal to a covariance matrix about $x_1$ and $x_2$. Thus, $\sigma^{-1}$ of the initial squeezed vacuum states, which we put $\hat{S}(r_1)\ket{0}\otimes \hat{S}(r_2)\ket{0}$, is given by
\begin{eqnarray}
\sigma^{-1} = \left(
    \begin{array}{cc}
      \Braket{\hat{x}_1^2} & \Braket{\hat{x}_1\hat{x}_2} \\
      \Braket{\hat{x}_2\hat{x}_1} & \Braket{\hat{x}_2^2}
    \end{array}
  \right) = 
 \left(
    \begin{array}{cc}
      \mathrm{e}^{-2r_1} & 0 \\
      0 & \mathrm{e}^{-2r_2}
    \end{array}
  \right).
\end{eqnarray}
Generally, beam splitters transform $\left(\hat{a}_1 \ \hat{a}_2\right)^T$ by an arbitrary unitary matrix $M_{{\rm BS}}$. In our case, we can assume $M_{{\rm BS}}$ is a real orthogonal matrix and $\bm{x}$ is transformed to $M_{{\rm BS}}\bm{x}$ because $\sigma$ is a real matrix. Then, the matrix $\sigma^{-1}$ is transformed to $M_{{\rm BS}}\sigma^{-1} M_{{\rm BS}}^{T}$. When the beam splitter has the power reflectance (transmittance) $R\ (T=1-R)$, $\sigma^{-1}$ and $\sigma$ are given by
\begin{eqnarray}
\sigma^{-1} &=& \left(
    \begin{array}{cc}
      \sqrt{R} & \sqrt{T} \\
      -\sqrt{T} & \sqrt{R}
    \end{array}
  \right)
\left(
    \begin{array}{cc}
      \mathrm{e}^{-2r_1} & 0 \\
      0 & \mathrm{e}^{-2r_2}
    \end{array}
  \right)
\left(
    \begin{array}{cc}
      \sqrt{R} & -\sqrt{T} \\
      \sqrt{T} & \sqrt{R}
    \end{array}
  \right) \nonumber \\
&=& \left(
    \begin{array}{cc}
      R\mathrm{e}^{-2r_1}+T\mathrm{e}^{-2r_2} & \sqrt{RT}\left( \mathrm{e}^{-2r_2}-\mathrm{e}^{-2r_1} \right) \\
      \sqrt{RT}\left( \mathrm{e}^{-2r_2}-\mathrm{e}^{-2r_1} \right) & T\mathrm{e}^{-2r_1}+R\mathrm{e}^{-2r_2}
    \end{array}
  \right),\label{eq:sigma-1}
\end{eqnarray}
\begin{eqnarray}
\sigma = \left(
    \begin{array}{cc}
      R\mathrm{e}^{2r_1}+T\mathrm{e}^{2r_2} & \sqrt{RT}\left( \mathrm{e}^{2r_1}-\mathrm{e}^{2r_2} \right) \\
      \sqrt{RT}\left( \mathrm{e}^{2r_1}-\mathrm{e}^{2r_2} \right) & T\mathrm{e}^{2r_1}+R\mathrm{e}^{2r_2}
    \end{array}
  \right).\label{eq:sigma}
\end{eqnarray}
Therefore, the conditions of GPS are given by
\begin{eqnarray}
\sigma_{11} &=& R\mathrm{e}^{2r_1}+T\mathrm{e}^{2r_2} = 1, \label{eq:cond1} \\
\sigma_{12} &=& \sqrt{RT}\left( \mathrm{e}^{2r_1}-\mathrm{e}^{2r_2} \right) \neq 0. \label{eq:cond2}
\end{eqnarray}
We can prepare desired $\ket{G}$ by selecting parameters $r_1,r_2 \in \mathbb{R}$ and $R\ (0\le R \le 1)$ satisfying these conditions. If and only if $r_1r_2<0$, there exists $R$ that satisfies Eqs.\ (\ref{eq:cond1}) and (\ref{eq:cond2}). Thus, the initial squeezed vacuum states should be squeezed in orthogonal directions. When $\sigma_{11}=1$, the squeezing factor of the outcome states in Eq.\ (\ref{eq:outcome}) is
\begin{eqnarray}
\mathrm{e}^{2r_c} = \mathrm{e}^{2(r_1+r_2)}+\mathrm{e}^{2r_1}+\mathrm{e}^{2r_2}-1.
\end{eqnarray}
Supposing $r_1=-r_2>0$ and $\mathrm{e}^{2r_1}\gg 1$, the generated cat states are as squeezed as the inputs because $\mathrm{e}^{2r_c} \approx \mathrm{e}^{2r_1}$.

%%%%%%%%%%%%%%%%%%%%%%%%%%%%%%%%%%%%%%%%%%%%%%%%%%%%%%%%%%%%%%%%%%%%%%%%%%%%%
\subsection{Dependence on $\sigma_{11}$\label{2d}}
%%%%%%%%%%%%%
\begin{figure*}[t]
	\begin{center}
		\includegraphics[bb= 0 0 860 280,clip,width=0.9\textwidth]{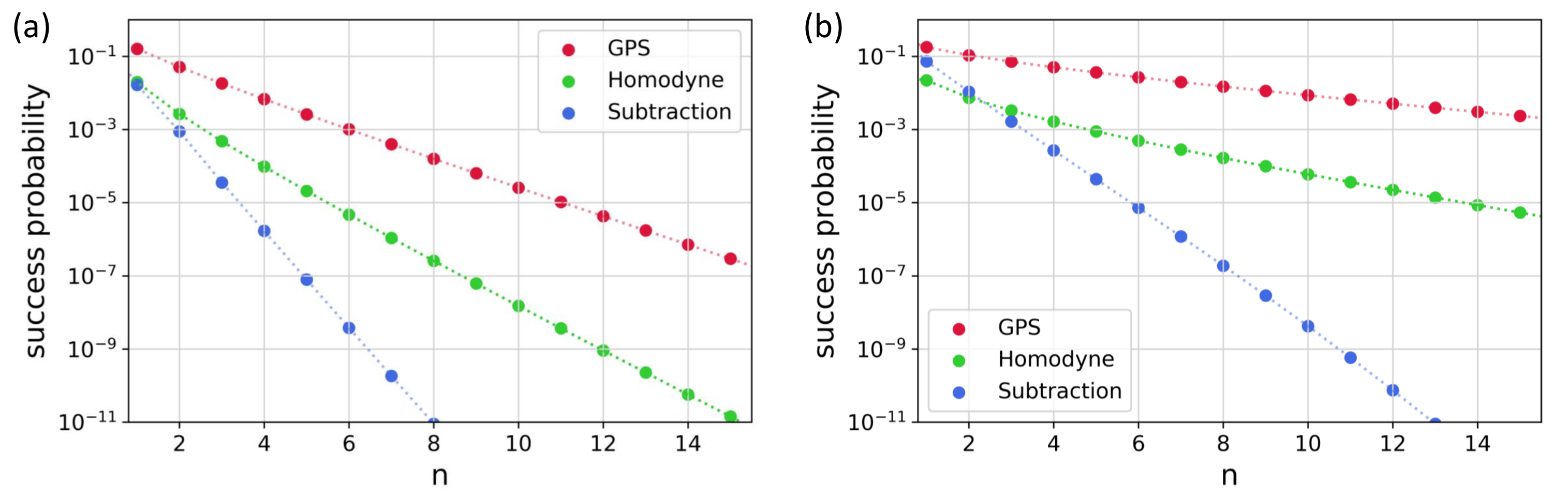}
	\end{center}
	\caption{(a),(b) Comparison of the success probability to generate a Schr\"{o}dinger's cat state with $\alpha = \sqrt{n}$. The compared methods are GPS, the homodyne conditioning method, and conventional photon subtraction. In each method, we suppose the squeezing parameters of inputs are $|r|=0.576$ (5 dB squeezing) in (a) and $|r|=1.15$ (10 dB squeezing) in (b).}
\label{Fig:main2}
\end{figure*}
%%%%%%%%%%%%
In this section, we discuss how the parameter $\sigma_{11}$ affects the outcome states and show that $\sigma_{11}=1$ is a reasonable condition. Figure\ \ref{Fig:main1}(a) shows the plots of the functions $\phi_n (x_1)$, $\phi_n (x_1) G (x_1,x_2)$, and $\psi_n(x_2)$ to visualize Eq.\ (\ref{eq:wavefunction}). From the left, each plot corresponds to $\sigma_{11} <1$, $\sigma_{11} =1$, and $\sigma_{11} >1$. We can see the tilted Gaussian structures of $G(x_1,x_2)$ striped by $\phi_n(x_1)$. In Fig.\ \ref{Fig:main1}(a), they get more tilted from $x_2$ axis as $\sigma_{11}$ increases. When $\sigma_{11}=1$, $\psi_n(x_2)$ has two peaks because the integral about $x_1$ averages out the stripe structure except for the two peaks on the both ends. $\psi_n(x_2)$ and the target cat state (black broken line) are almost identical. When $\sigma_{11}<1$, $\psi_n(x_2)$ has a cat-like waveform but its amplitude decreases. When $\sigma_{11}>1$, the interval of the highest peaks increases but an unwanted oscillation appears. This is because the Gaussian shape tilts too much for the stripe to cancel. In an analytical perspective, we can derive $\Psi_n(x_2)$ from Eqs.\ (\ref{eq:wavefunction}) and (\ref{eq:g}) as follows,
\begin{eqnarray}
\Psi_n(x_2) &=& |\sigma |^{\frac{1}{4}}\ 
\phi_0 \left( \sqrt{\frac{\left| \sigma \right|}{\sigma_{11}}} x_2 \right) I_n^{(\sigma_{11})}\left( -\frac{\sigma_{12}}{\sigma_{11}}x_2 \right), \\
I_n^{(\sigma_{11})}(x) &=& \int dy \ \phi_0\left(\sqrt{\sigma_{11}}(x-y)\right)\phi_n(y).
\end{eqnarray}
The waveform of $\psi_n(x_2)$ is mainly decided by the function $I_n^{(\sigma_{11})}(x)$. When $\sigma_{11}=1$, $\psi_n(x_2)$ is close to the wavefunction of cat states due to the relation $I_n^{(1)}(x)=\left( \phi_0*\phi_n \right)(x)$. When $0<\sigma_{11} < 1$, we can derive
\begin{eqnarray}
I_n^{(\sigma_{11})}(x) = \left( g * I_n^{(1)} \right) (x)\ , \ g(x) = \frac{\exp{\left(-\frac{\sigma_{11}}{2(1-\sigma_{11})}x^2 \right)}}{\sqrt{2(1-\sigma_{11})\pi}}.
\end{eqnarray}
Thus, we have another Gaussian convolution on $ I_n^{(1)}(x)$. In this case, we still have a cat-like wavefunction, but its effective amplitude decreases due to the extra convolution. When $\sigma_{11} > 1$, $\phi_n(x)$ is convolved by a Gaussian function narrower than $\phi_0(x)$. In this case, an unwanted oscillation remains in $\psi_n(x_2)$ because the oscillation of $\phi_n(x)$ is not averaged out completely.

Figure\ \ref{Fig:main1}(b) shows the Fourier counterpart of the functions in Fig.\ \ref{Fig:main1}(a), that is, $\tilde{\phi}_n(p_1)$, $\tilde{\phi}_n(p_1)\tilde{G}(p_1,p_2)$, and $\tilde{\psi}_n(p_2)$. The functions $\phi_n(x_1)$ and $\tilde{\phi}_n(p_1)$ have the same waveform because $\ket{n}$ is phase insensitive. $\tilde{G}(p_1,p_2)$ is characterized by a matrix $\tilde{\sigma}$, which is equal to $\sigma$ with the sign inversion of $r_1,r_2$. From Eqs.\ (\ref{eq:sigma-1}) and (\ref{eq:sigma}), the covariance matrix of $\ket{G}$ about $p_1,p_2$ is given by
\begin{eqnarray}
\tilde{\sigma}^{-1} = \left(
    \begin{array}{cc}
      \Braket{\hat{p}_1^2} & \Braket{\hat{p}_1\hat{p}_2} \\
      \Braket{\hat{p}_2\hat{p}_1} & \Braket{\hat{p}_2^2}
    \end{array}
  \right) = 
 \left(
    \begin{array}{cc}
      \sigma_{11} & -\sigma_{12} \\
      -\sigma_{21} & \sigma_{22}
    \end{array}
  \right). \label{eq:tildesigma}
\end{eqnarray}
From Eq.\ (\ref{eq:wavep}), $\tilde{\psi}_n(p_2)$ should have cosine (or sine) oscillations with a Gaussian envelope. Wentzel-Kramers-Brillouin approximation \cite{Schleich2005} shows $\tilde{\phi}_n(p_1)$ has cosine (or sine) oscillations when $p_1$ is small. In GPS, these oscillations of $\tilde{\phi}_n(p_1)$ are mapped to $\tilde{\psi}_n(p_2)$ by a Gaussian function $\tilde{G}(p_1,p_2)$. A wider range of $\tilde{\phi}_n(p_1)$ structure appears in $\tilde{\psi}_n(p_2)$ as the variance of $\ket{G}$ about $p_1$ increases. Thus, $\Braket{\hat{p}_1^2}=\sigma_{11}$ has a critical effect on the waveform of $\tilde{\psi}_n(p_2)$. When $\sigma_{11}=1$, $\tilde{\psi}_n(p_2)$ well approximates the ideal line. When $\sigma_{11}<1$, smaller number of cosine oscillations appear in $\tilde{\psi}_n(p_2)$. That means the amplitude of the generated cat state gets smaller. When $\sigma_{11}>1$, un-cosinusoidal structure of $\tilde{\phi}_n(p_1)$ is mapped to $\tilde{\psi}_n(p_2)$, which makes the generated states away from ideal cat states.

Like the above, the two distinct areas of the wavefunction of $\ket{n}$, two peaks and cosinusoidal oscillations, appear in $\psi_n(x_2)$ and $\tilde{\psi}_n(p_2)$ through the Gaussian functions $G(x_1,x_2)$ and $\tilde{G}(p_1,p_2)$, respectively. Supposing $\sigma_{11}=1$, we can ensure that high-fidelity and large-amplitude cat states are generated.

%%%%%%%%%%%%%%%%%%%%%%%%%%%%%%%%%%%%%%%%%%%%%%%%%%%%%%%%%%%%%%%%%%%%%%%%%%%%%
\section{Evaluation of generation rate\label{3}}
GPS can generate cat states at a much better rate than conventional methods. From Eqs.\ (\ref{eq:pn}) and (\ref{eq:unnormwf}), the probability to obtain $\ket{\psi_n}$ in the condition $\sigma_{11}=1$ is
\begin{eqnarray}
P(n) = \frac{\sqrt{|\sigma|}(2n)!(\sigma_{12})^{2n}}{8^n(n!)^2} \left( \frac{\left| \sigma \right|+\sigma_{22}}{2} \right)^{-n-\frac{1}{2}}.
\end{eqnarray}
GPS contains some previous works as special cases due to its generality. In these works, $P(n)$ was quite low because $|\sigma_{12}| = \sqrt{RT}\left| \mathrm{e}^{2r_1}-\mathrm{e}^{2r_2} \right| \ll 1$ is assumed. For example, conventional photon subtraction assumes $R \ll 1$ and $r_2=0$. The weak tapping condition $R \ll 1$ makes it difficult to detect photons in the tapped mode. In another example \cite{Huang2015}, two squeezed vacuum states are utilized but the low input power condition $r_1=-r_2\ll 1$ is assumed. Now, we have a generalized condition for cat-state generation $\sigma_{11}=1$, so that we can select parameters that avoid undesirable conditions like $R\ll 1$ or $|r_1 -r_2| \ll 1$. In addition, GPS performs conditioning only once, and thus it is advantageous than other methods that perform conditioning more than once \cite{Ourjoumtsev2007,Ulanov2016,Sychev2017}. Those factors indicate the potential of GPS for improvement of the state generation rate.

We compare the cat-state generation rate of GPS with the homodyne conditioning method \cite{Ourjoumtsev2007} and conventional photon subtraction. In GPS, we assume $r_1=-r_2>0$ and select $R$ that satisfies $\sigma_{11}=1$.  In the homodyne conditioning method, a Fock state $\ket{n}$ is generated from a two-mode squeezed vacuum state and $n$ photon detection, followed by 50:50 beam splitting and homodyne conditioning in one mode. When we use squeezed vacuums $\hat{S}(r)\ket{0}$ as inputs and generate cat states with fidelity about 0.99, the success probability of $n$ photon detection and homodyne conditioning are $(1-\tanh^2{r})\tanh^{2n}{r}$ and $1/(10n)$, respectively. In conventional photon subtraction, we assume $R=0.05$ and numerically calculate the success probability in a subspace up to 50 photons by a Python library for photonic quantum computing \cite{Killoran2019strawberryfields,Bromley2020}. 

Figures\ \ref{Fig:main2}(a) and \ref{Fig:main2}(b) are the success probability to generate the cat states $\Ket{{\rm Cat}_{\sqrt{n},n}}$ with squeezing. We assume that the squeezing parameters of inputs are $|r|=0.576$ (5 dB squeezing) in Fig.\ \ref{Fig:main2}(a), and $|r|=1.15$ (10 dB squeezing) in Fig.\ \ref{Fig:main2}(b) in each method. In the both cases, GPS has the highest success probability, and the superiority increases as $n$ increases. Especially, the improvement from conventional photon subtraction is remarkable. The improvement of the success rate easily reaches several orders as $n$ increases. GPS is also better than the homodyne conditioning method by multiple orders. In this case, the difference of success rates mainly comes from the number of conditioning. The success rate to generate a cat state in GPS and the rate to generate a Fock state in the homodyne conditioning method are comparable, but the need of one more conditioning process in the latter method than the former one lowers the total success rate.

%%%%%%%%%%%%%
\begin{figure}[t]
	\begin{center}
		\includegraphics[bb= -20 0 740 260,clip,width=0.8\textwidth]{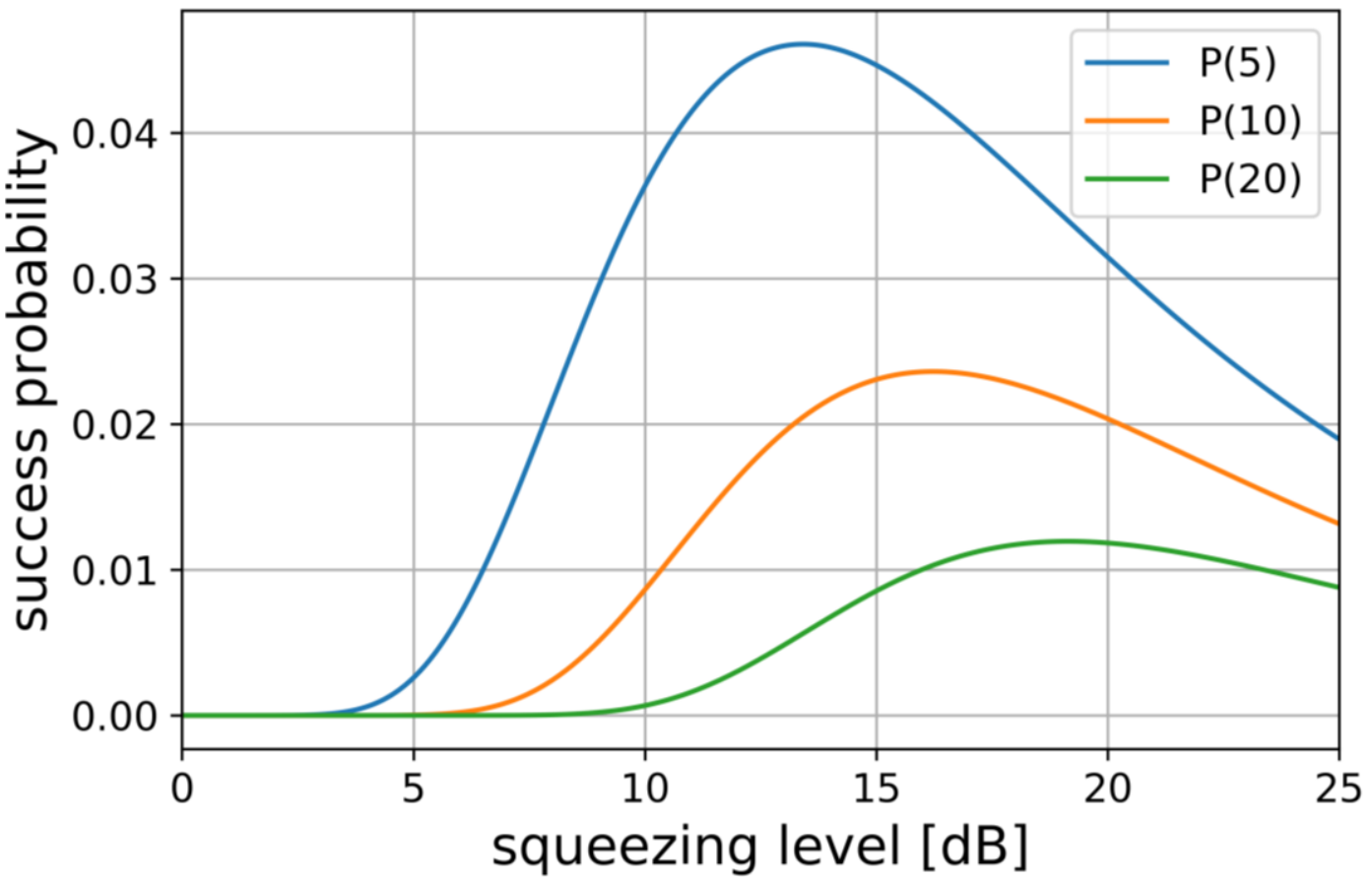}
	\end{center}
	\caption{The probability to detect 5, 10, and 20 photons in GPS in the case of $r_1=-r_2$ and $\sigma_{11}=1$.}
\label{Fig:main3}
\end{figure}
%%%%%%%%%%%%
Finally, we show a generation rate estimation of GPS. The cat states with $\alpha \ge 2$ are desired in quantum computing \cite{Ralph2003}. Thus, we are interested in the case of $n\ge 4$. Figure\ \ref{Fig:main3} is the behaviors of $P(5),P(10),$ and $P(20)$ against the input squeezing level on the assumption of $r_1=-r_2$ and $\sigma_{11}=1$. We see they have each maximum value at some points. This is because the distribution of $P(n)$ gradually becomes flat as the squeezing level increases in the condition of $\sum P(n) =1$. Thus far, $r=1.73$ (15 dB squeezing) has been demonstrated \cite{Vahlbruch2016}. Supposing $r_1=-r_2=1.73$, we get $P(10)=0.023$. The generation rate of $\ket{\psi_{10}}$ becomes $0.023\times f_{{\rm rep}}$ by operating the system at a rate $f_{{\rm rep}}$. The performances of squeezed vacuum sources and PNRDs decide the limit of $f_{{\rm rep}}$. Recent works argue that implementation of PNRDs by multiplexed on-off detectors is demanding \cite{Joensson2019,Provaznik2020}, and thus other methods like transition edge sensors or superconducting nanowire detectors are desired \cite{Fukuda2011,Lita2008,Nehra2019,Sridhar2014,Cahall2017}. Because the experimental results so far \cite{Cahall2017,Takanashi2019,Kashiwazaki2020} indicate that $f_{{\rm rep}}=100$ MHz is possible, we have enough chance to generate $\ket{\psi_{10}}$ at Mcps order. Refining the performances of squeezed vacuum sources and PNRDs leads to the further improvement of this rate. Since current optical CV information processors work at MHz order \cite{Asavanant2020}, single cat-state source of GPS might be enough to feed cat states into the processor as inputs. This rate is $10^3$ to $10^6$ times better than conventional photon subtraction where we assume $0.02\le R \le 0.05$ as a typical condition. Like the above, GPS would lead to generation of the large-amplitude cat states at the rate enough for implementation of quantum optical applications.

%%%%%%%%%%%%%%%%% END OF THEORY %%%%%%%%%%%%%%%%%%%%%%%%%%%%%%%%%%%%%%%%%%
\section{Conclusion\label{4}}
We have proposed GPS for generation of optical Schr\"{o}dinger's cat states. We started from a generalized situation of photon number measurement on a arbitrary two-mode Gaussian state, and derived the conditions of cat-state generation analytically. Our method relaxes the constraints on experimental parameters compared to conventional methods, allowing us to select optimal parameters and improve the generation rate by multiple orders. Supposing realistic experimental conditions, the generation rate of the large-amplitude cat states ($\alpha \ge 2$) is expected to reach Mcps order, which is as fast as the system clock of current CV quantum information processors. Because the performance of GPS is limited by light sources and PNRDs, the generation rate would be much faster than Mcps order by the progress of these factors. GPS is feasible in free space thanks to its simple setup. Each component of GPS has been implemented on a chip \cite{Montaut2017,Kashiwazaki2020,Masada2015,Lenzini2018}, and thus the integration of our cat-state sources would be possible in the future. Our proposal is important in optical CV quantum computing, where information processing platforms are ready but high-rate supply of input non-Gaussian states remains as a challenging task. Our method would reduce the difficulties in the state generation system remarkably, and make a significant progress toward fault-tolerant CV quantum computing.

\section{Acknowledgements}
This work was partly supported by JSPS KAKENHI (Grant No. 18H05207, No. 18H01149, and No. 20K15187), the Core Research for Evolutional Science and Technology (CREST) (Grant No. JPMJCR15N5) of the Japan Science and Technology
Agency (JST), UTokyo Foundation, and donations from Nichia Corporation.
K. T. and W. A. acknowledge financial supports from the Japan Society for the Promotion of Science (JSPS). The authors would like to thank Takahiro Mitani for careful proofreading of the manuscript.

%%%%%%%%%%%%%%%%%%%%%%%%%%%%%%%%%%%%%%%%%%%%%%%%%%%%%%%%%%%%%%%%%%%%%%%%%%%%%%%%%%%%%%%%%%%%%%%%%%%%%%%%%%%%

%\bibliographystyle{unsrt}

%\bibliography{scibib}

\end{document}